\documentclass[twocolumn,showpacs,aps]{revtex4}
\usepackage{amssymb}
\usepackage{amsmath}
\usepackage{graphicx}
\usepackage{bm}
\usepackage{epstopdf}
\usepackage[utf8x]{inputenc}
\usepackage{hyperref}
\usepackage[all]{hypcap}
\hypersetup{
    colorlinks=true,
    linkcolor=blue,
    filecolor=magenta,
    urlcolor=cyan,
    citecolor=red
     }
\begin{document}

\title{Sub-femtosecond wakefield injector and accelerator based on an undulating plasma bubble controlled by a laser phase}

\author{Jihoon Kim$^1$, Tianhong Wang$^1$,  Vladimir Khudik$^1$$^,$$^2$, and Gennady Shvets$^1$}
\affiliation{$^1$School of Applied and Engineering Physics, Cornell University, Ithaca, NY 14850, USA.\\$^2$Department of Physics and Institute for Fusion Studies, The University of Texas at Austin, Austin, TX 78712, USA.
}

\begin{abstract}

We demonstrate that a long-propagating plasma bubble executing undulatory motion can be produced in the wake of two co-propagating laser pulses: a near-single-cycle injector and a multi-cycle driver. When the undulation amplitude exceeds the analytically-derived threshold, highly-localized injections of plasma electrons into the bubble are followed by their long-distance acceleration. While the locations of the injection regions are controlled by the carrier-envelope phase (CEP) of the injector pulse, the mono-energetic spectrum of the accelerated sub-femtosecond high-charge electron bunches is shown to be nearly CEP-independent.

\end{abstract}

\maketitle

Laser-driven plasma accelerators offer a promising pathway to compact accelerators by sustaining electric fields capable of accelerating charged particle to GeV energies in less than a centimeter. In addition, the plasma medium can serve as a cathode by supplying the electrodes to be trapped and accelerated inside a plasma cavity generated in the wake of an ultra-intense laser pulse via time-averaged (ponderomotive) pressure. Such laser-wakefield accelerators(LWFA)~\cite{Malka,RMPS,Hooker} have produced multi-GeV, low-emittance, ultra-short electron bunches \cite{Nakamura, Wang, Leemans, Kim, Gonsalves} without a need for a separate cathode.  High-energy electrons generated by LWFAs are promising candidates for various scientific and technological applications: from TeV-scale lepton colliders~\cite{TeV} to novel sources of high brightness radiation and particles~\cite{Park,Kneip,Stark,Schollmeier,Pomerantz,Chen}.

Under most circumstances relevant to laser-plasma accelerators, phase-averaged (ponderomotive) description of the plasma response to multi-cycle laser pulses~\cite{Mora} is sufficient for explaining the key phenomena enabling the LWFAs. Those include relativistic self-guiding responsible for the LWFA lengths much exceeding the vacuum diffraction distance of a laser pulse~\cite{RMPS}, as well as the dynamics of the plasma bubble determining electron acceleration and injection~\cite{kalmykov,gonsalves_nphys11,austin_injection}. However, the absolute carrier-envelope phase (CEP) of near-single-cycle (NSC) laser pulses can be important for some applications, including ionization injection~\cite{chen_jap06,pak_prl10,McGuffey_prl10,mori_bunches_prl16} by ultra-short laser pulses~\cite{lifschitz_malka_njp12,ouille_lsa20}. Combined electron injection and acceleration by NSC pulses has also been proposed~\cite{kost_cep,ma_screp16,CEP_observable,Zhengyan} and experimentally demonstrated~\cite{kHz_injection,Veisz_2cycle,Salehi} in fully-ionized plasmas. However, significant shot-to-shot variation of the electron energy spectrum due to CEP slip has been observed~\cite{Salehi} for NSC pulses without CEP stabilization. Overall, NSC laser pulses have limited potential as drivers for GeV-scale LWFAs because their reduced self-focusing~\cite{Sprangle_guide} results in rapid diffraction and short acceleration distance~\cite{kHz_injection,CEP_observable,Zhengyan}.

\begin{figure}[t]
    \includegraphics[width=0.5\textwidth]{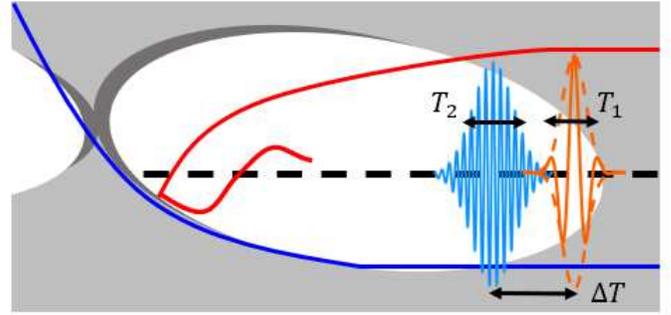}
\caption{Schematic of a laser-wakefield accelerator with phase-controlled undulating bubble (PUB) injection. A transversely undulating plasma bubble driven by the combination of a near-single-cycle injector pulse (orange) and a multi-cycle driver pulse (light blue) periodically traps electrons from the ambient plasma. Time-dependent injector CEP controls bubble centroid displacement from the propagation axis (dashed line), determines which electrons are injected into (red line) or pass through (blue line) the bubble. }\label{fig:schematic}
\end{figure}

In this Letter, we show that an NSC laser pulse can be harnessed as an ultra-fast electron injector when combined with a multi-cycle higher-intensity laser pulse serving as a LWFA driver over long distances. Using 3D Particle-In-Cell (PIC) simulations, we show that asymmetric plasma flow controlled by the CEP of a moderate-power injector pulse induces undulations of a long-lived plasma bubble generated by the driver pulse. When the undulation amplitude exceeds the analytically derived threshold, electron injection by a phase-controlled undulating bubble (PUB) takes place. This two-pulse approach shown in Fig.~\ref{fig:schematic} turns the key disadvantage of NSC pulses -- their short propagation distance in tenuous plasmas -- into an advantage: highly localized electron injection. Because almost all of the energy gain takes place after the depletion of the injection pulse, PUB-injected electrons form sub-fs high-current ($10'$s of kAs) mono-energetic electron bunches with CEP-independent energy spectrum.

{\it Analytic criterion for electron trapping by an undulating bubble.} We use a simplified model of a positively-charged (devoid of electrons) spherical plasma bubble with radius $R$  propagating with uniform velocity $v_b$~\cite{kostyukov_pop04,kalmykov}. A moving-frame Hamiltonian describing plasma electrons' interaction with the bubble is given by $H(\rho,t) = \sqrt{1+(\mathbf{P}+\mathbf{A})^2} - v_b P_x - \phi $~\cite{kost_injection,kalmykov,austin_injection}, where $\mathbf{\rho} = (\xi,y,z-z_{\rm osc})$, $\xi=x-v_b t$, $z_{\rm osc}(t)$ is the transverse coordinate of the undulating bubble center, $\mathbf{P}$ is the canonical momentum, and $\mathbf{A}$ ($\phi$) are the vector (scalar) potentials. Time, length, potential, and electron momentum are normalized to $\omega_p^{-1}$, $k_p^{-1} = c/\omega_p$, $m_e c^2/|e|$, and $m_e c$, respectively, where $\omega_p=\sqrt{4\pi e^2 n_p/m}$ is the electron plasma frequency and $n_p$ is the plasma density.

Under the $A_x = -\phi = \Phi/2$  gauge, we further assume that $\Phi=(\rho^2-R^2)/4$ inside and $\Phi=0$ outside the bubble~\cite{kost_injection}. Transverse undulations $z_{\rm osc}(t) \equiv z_{\rm u}\cos(\omega_{\rm CEP} t + \phi_{\rm CEP})$ of the plasma bubble excited by the injector pulse introduce time dependence into the Hamiltonian. Here $\omega_{\rm CEP}\equiv 2\pi/T_{\rm CEP}$ is the injector CEP slip rate with respect to the bubble speed~\cite{kost_cep,ma_screp16,CEP_observable,Zhengyan,Salehi}, $z_{u}$ is the maximum bubble oscillation amplitude, and $\phi_{\rm CEP}\equiv \phi_{\rm CEP}(t(x_0),x_0)$ is the initial CEP evaluated at the time $t(x_0)$ corresponding to electron's entrance into the bubble at $x=x_0$. Under the relativistic approximation $v_b/c \approx 1-1/2\gamma_b^2$ for $\gamma_b \gg 1$, where $\gamma_b$ is the relativistic factor of the bubble, the undulation period is
\begin{equation}\label{eq:period_cep}
  cT_{\rm CEP} \approx 2 \lambda_{\rm inj} \left( \frac{1}{\gamma_b^2} + \frac{n_p}{n_{\rm crit}(\lambda_{\rm inj})} \right)^{-1},
\end{equation}
where $n_{\rm crit}(\lambda_{\rm inj})=\pi m_e c^2/(e^2\lambda_{\rm inj}^2)$ is the critical density for the injector wavelength $\lambda_{\rm inj}$.

Electron equations of motion in the $x-z$ plane derived from $H(\rho,t)$ are:
\begin{flalign}
&\frac{d\xi}{dt} = \frac{p_x}{\gamma} - v_b,  &&\frac{d p_x}{dt} =-\frac{1}{4} \left( \xi(1+v_b) + (v_z - \dot{z}_{\rm osc}) \Tilde{z}\right)& \label{eq:EqM_N1} \\
&\frac{dz}{dt} = \frac{p_z}{\gamma},  &&\frac{d p_{z}}{dt} = -\frac{(v_x+1)\Tilde{z}}{4} \nonumber
\end{flalign}
where the explicitly time-dependent terms $z_{\rm osc}$ and $\tilde{z}(t)= z-z_{\rm osc}$ change the Hamiltonian from its initial value of $H(t)=1$ for the quiescent electrons in front of the bubble. Bubble undulations cause the Hamiltonian to evolve according to $dH/dt=\partial H/\partial t$, enabling some of the plasma electrons to get trapped inside the bubble when the following condition is satisfied:
\begin{equation}
    \Delta H = \int dt \dot{p_z}(t)\dot{z}_{\rm osc}(t) < -1,
    \label{eq:EqM_N5}
\end{equation}
where the integral is calculated along the electron trajectory~\cite{kalmykov,austin_injection}.


To lowest order in the bubble undulation amplitude $z_{\rm u}$, and assuming that electron passage time through the bubble $T_{\rm pass}\sim R$ is much shorter than $T_{\rm CEP}$, the Hamiltonian increment can be approximated as $\Delta H^{(1)} \approx - z_{\rm u}\omega_{\rm CEP} \sin(\phi_{\rm CEP}) \Delta p^{(0)}$, where $\Delta p^{(0)}$ is the zeroth-order ($z_{\rm u}=0$) transverse momentum change of an electron passing through the bubble. For an electron entering the bubble at its edge at $z=\pm R$ and pulled inside, the maximum transverse momentum change is $|p_z^{\rm max}| \approx 0.14 R^2$ in the limit $v_b=c$ (See Fig.S1 of SOM). Therefore, the trapping condition for an electron entering the bubble's edge at the optimal phase ($\phi_{\rm CEP}=\pm \pi/2$) is estimated as $z_{\rm u} >7/(\omega_{\rm CEP}R^2)$ in normalized units. Assuming $\omega_{\rm CEP} \sim \lambda_{\rm inj}\omega_p^2/\left( 2\pi c\right)$~\cite{kost_cep}, the injection criterion is expressed as $z_{\rm u} > z_{\rm u}^{\rm tr}$, where the trapping threshold is $z_{\rm u}^{\rm tr}/R \sim 7 \left(k_p R \right)^{-3} \sqrt{n_{\rm crit}/n_p}$ in physical units.

\begin{figure}[t]
    \includegraphics[width=0.5\textwidth]{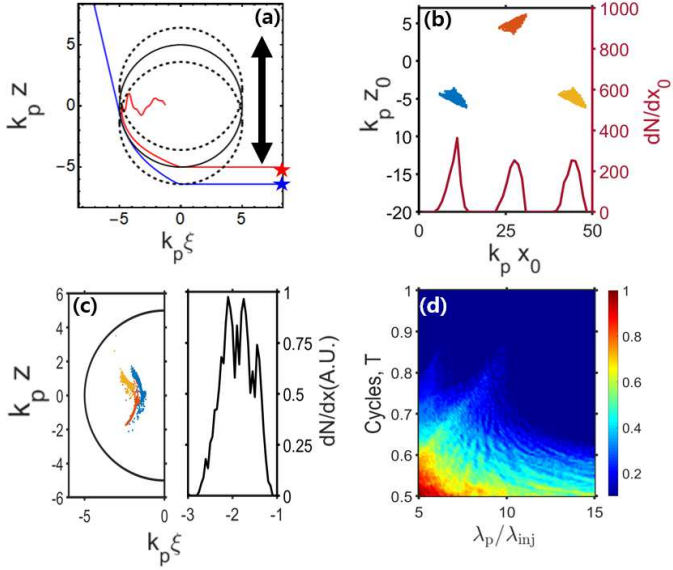}
\caption{
(a-c) Trapping of test particles by an undulating bubble and (d) bubble asymmetry created by an NSC pulse. (a) Trapped (red: $\phi_{\rm CEP}=\pi/2$) and passing (blue: $\phi_{\rm CEP}=\pi$) trajectories. Bubble boundaries: unperturbed (black solid line) and maximally-displaced (black dashed line). (b-c) Simulation of a particle swarm with a range of initial conditions $(x_0,y_0,z_0)$ color-coded by $x_0$. Only electrons trapped in the bubble at $t=200$ are plotted in the $(x_0,z_0)$ plane (b) and inside the bubble (c). Red line in (b): density of injected electrons vs their initial position. Black line in (c): longitudinal density distribution. PUB parameters: $k_p R=5$, $z_{\rm u}/R=0.28$, $T_{\rm CEP}=35/\omega_p$, $\gamma_b= 5$. (d) Plasma flow asymmetry: $\delta z_{\rm ex}$ (color-coded) of the electrons after passing through the injector pulse and non-oscillating ($z_{\rm u}=0$) bubble v.s. NSC wavelength $\lambda_{\rm inj}$ and pulse length $T$.}\label{fig:trajectory}
\end{figure}

\begin{figure}[t]
\centering
    \includegraphics[width=0.5\textwidth]{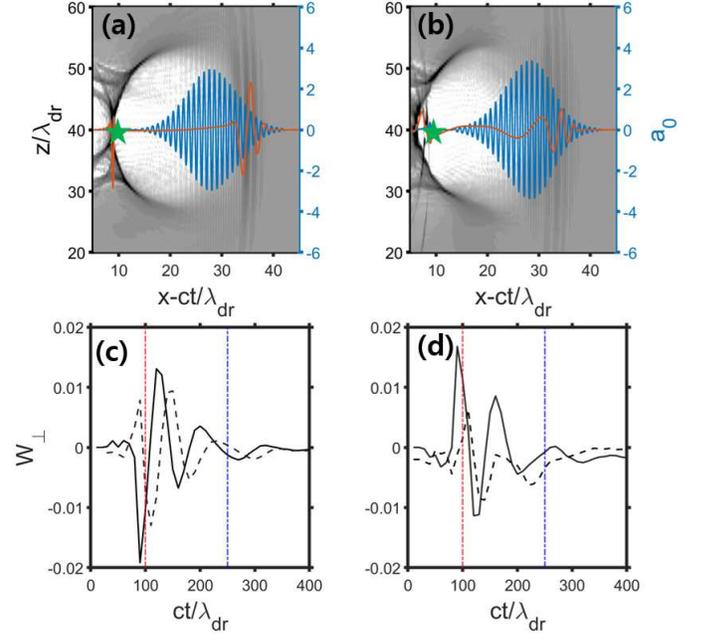}
\caption{Evolution of the injector/driver pulses and transverse plasma wakes. (a-b) Grayscale: plasma density in the $x-z$ plane around the bubble. Red (blue) lines: on-axis normalized electric fields of the injector (driver) pulses at  (a) $x_1=0.08 {\rm mm}$ and (b) $x_2=0.2 {\rm mm}$.  (c-d) On-axis transverse wakes at $\zeta \equiv (x-ct)/\lambda = 9$ (green stars in (a-b)). (c) CEP dependence of $W_{\perp}$ for $\lambda_{\rm inj}=2.4\mu m $ injector pulse: $\phi_{\rm CEP}=0$ (black solid line) and $\phi_{\rm CEP}=\pi/2$ (black dashed line). (d) Dependence of $W_{\perp}$ on the injector pulse wavelength: $\lambda_{\rm inj}=2.4\mu m $ (black solid line) and $\lambda_{\rm inj}^{(2)}=1.2\mu m $ (black dashed line); $\phi_{\rm CEP}=\pi$ for both wavelengths. Red (blue) dotted-dashed lines: propagation distances $x_1$ ($x_2$). All fields scaled to $E_0=e/mc\omega_{\rm dr}$. Plasma density: $n_p=4.4\times 10^{18}/cm^3$, laser parameters: see Table I. Simulations: 3D PIC VLPL.}\label{fig:evolution}
\end{figure}

\begin{figure}[t]
\centering
    \includegraphics[width=0.5\textwidth]{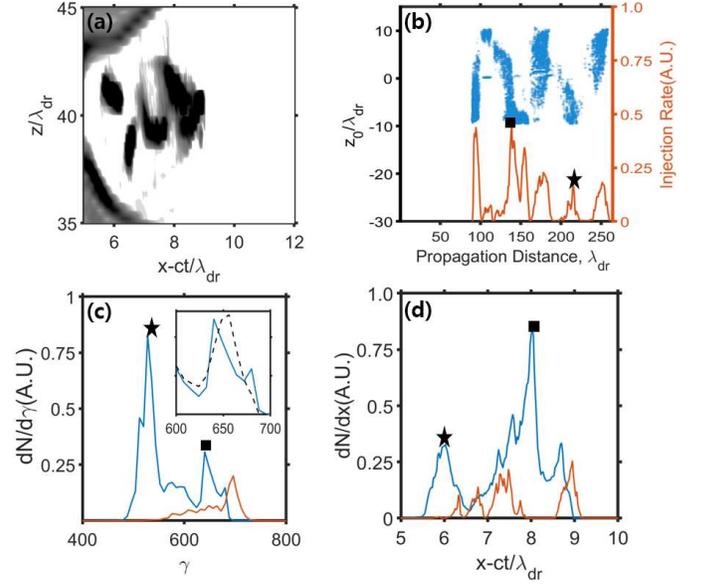}
\caption{PUB-based injection and acceleration. (a) Grayscale: density crossection at $x=0.32 {\rm mm}$. (b) Electron injection rate (solid line) and injected electrons' initial transverse positions (blue dots) as a function of propagation distance.
(c-d) Injected electrons' energy spectra (c) and current (d) at $x=1.5 {\rm mm}$ for Injector 1 (blue line) and Injector 2 (red line). Star and square in (b-d): the corresponding injection times (b), energy spectra (c), and positions inside the bubble (d). Inset in (c): high-energy spectral peaks at $\phi_{\rm CEP} = 0$ (solid line) and $\phi_{\rm CEP} = \pi/2$ (dashed line). Laser parameters: see Table I.
Simulations: 3D PIC VLPL}\label{fig:injected}
\end{figure}

PUB-based injection is visualized in Fig.~\ref{fig:trajectory}(a), where the trajectories given by Eqs.(\ref{eq:EqM_N1}) are plotted for two initially quiescent electrons. The first (red star: trapped) electron enters the undulating bubble with $\phi_{\rm CEP}=\pi/2$, while the second (blue star: passing) one is delayed in time, entering the bubble with $\phi_{\rm CEP}=\pi$. The electrons' initial transverse positions (red star:$z_0=-R$, blue star: $z_0=-(R+z_u)$) are chosen such that electrons enter the bubble at its lower edge, and the undulating bubble parameters (see caption) approximately correspond to those of the 3D PIC simulation presented later (see Figs.~\ref{fig:evolution},\ref{fig:injected}). The trapped electron's Hamiltonian $H(t=15)<0$  fulfils the trapping condition $\Delta H < -1$, while the passing electron's Hamiltonian $H(t)>0$ increases and remains positive. Moreover, the trapping fraction of the particles incident onto the bubble rapidly vanishes for $z_{\rm u} < z_{\rm u}^{\rm tr}$ (see Fig.S1(f) of the SOM).
As this calculation demonstrates, injections occur twice per period: if a bubble traps electron at $(x_1,z_1)$ for $\phi_{\rm CEP1}$, then  it will also trap a ``partner" electron for $\phi_{\rm CEP2} = \phi_{\rm CEP1} + \pi$ at $(x_2=x_1 + v_b T_{\rm CEP}/2,z_2=-z_1)$.

To model electron trapping from background plasma, we next simulate the interaction of the undulating bubble with a swarm of initially resting electrons randomly seeded into a 3D volume of initial positions (longitudinal $R<x_0<50$ and transverse $-6.5 < y_0,z_0 < 6.5$) entering the bubble during the $0<t<200$ time interval. Electron injection occurs every half-period of the bubble oscillation as shown in Fig.~\ref{fig:trajectory}(c), where electrons are color-coded based on their longitudinal injection location $x_0$ (or, equivalently, injection time $t_0=x_0/v_b$). Note that the injected electrons primarily originate from the bubble's edge: since $|\Delta H| \propto \bar{\rho}^2$ (where $\bar{\rho}$ is the impact parameter of an electron entering the bubble), electrons grazing the bubble at $\bar{\rho} \approx R$ are the best candidates for trapping.

Time delay between separate injections determines the longitudinal structure of the trapped/accelerated electrons inside the bubble. As electrons rapidly accelerate, they become ultra-relativistic and advance through the bubble. Therefore, longitudinal spacing between micro-bunches entering the bubble at the adjacent injection times can be written as $\Delta \xi \approx (c-v_b) \times T_{\rm CEP}/2  \approx T_{\rm CEP}/(4 \gamma _b^2)$. The resulting texture of the modulated injected beam is shown in Figure \ref{fig:trajectory}(c), where several ultra-short bunches correspond to different (color-coded) PUB-controlled injection times.

{\it Inducing bubble undulations by an NSC pulse.} Next, using single-particle simulations, we establish the optimal wavelength $\lambda_{\rm inj}$ and duration $\sigma_x \equiv T \lambda_{\rm inj}$ of the injector pulse producing the largest asymmetric plasma flow around the fixed bubble. Flow asymmetry can be used as a proxy for induced bubble undulation amplitude $z_u$. We use particle swarm simulations similar to those used in Figs.~\ref{fig:trajectory}(b,c), except that the bubble is assumed to be non-undulating, and the electric field of the injector pulse placed ahead of the bubble is given by
$E_z=a_{\rm inj} \omega_{\rm inj}e^{-(y^2+z^2)/\sigma_{\rm inj}^2}e^{-\left( \xi - R\right)^2/\sigma_x^2}  \times \cos{\left[ \omega_{\rm inj}\left( x - v_{\rm ph}t -R\right) + \phi_{\rm CEP}\right]}$.
Here the fixed parameters are the injector pulse spot size $\sigma_{\rm inj}=3$ and vector potential $a_{\rm inj}=4$, while the cycle number $T$ and the normalized injector pulse frequency $\omega_{\rm inj}/\omega_{\rm p} = \lambda_{\rm p}/\lambda_{\rm inj}$ are varied [See SOM for details.] Each particle is removed from the simulation after reaching the end of the bubble at the exit time $t_{\rm ex}$ such that $\xi(t_{\rm ex}) = -R$, and its transverse displacement is $z_{\rm ex} \equiv z(t_{\rm ex})$.

The amplitude of the particle-averaged exit transverse displacement asymmetry $\delta z_{\rm ex} \equiv \left< z_{\rm ex} \right>$ plotted in Fig.~\ref{fig:trajectory}(d) is enhanced for (1) longer $\lambda_{\rm inj}$ and (2) shorter $T$. While this finding is not surprising in the light of the non-ponderomotive scaling $\delta p_z \propto a_0^{3} \lambda_{\rm inj}^2 \sin(\phi_{\rm CEP})/\sigma_{\rm inj}^2 T^2$~\cite{kost_cep} of the momentum asymmetry, we note that the latter analytic expression has been derived in the absence of the plasma bubble. Taking into account the injection threshold scaling $z_{\rm u}^{\rm tr} \propto 1/\lambda_{\rm inj}$, it follows that a long-wavelength NSC injector is advantageous for particle trapping.

\

{\it PIC simulations.} The single-particle model suggests the following sequence of events when an NSC injector pulse co-propagates with a strong driver pulse: (i) asymmetric plasma flow around the combination the injector pulse and plasma bubble generated by the driver pulse produces an undulating bubble ( Fig.~\ref{fig:trajectory}(d)), (ii) the latter periodically traps plasma electrons (Fig.~\ref{fig:trajectory}(b)), and (iii) produces a structured bunch (Fig.~\ref{fig:trajectory}(c)). We use a 3D PIC code VLPL\cite{Pukhov_code} to self-consistently model multiple physical effects accompanying nonlinear interactions between the two laser pulses and the plasma, including: laser self-guiding~\cite{Sprangle_guide}, depletion of the plasma fields by injected electrons~\cite{beamloading}, the  non-spherical structure of the plasma bubble\cite{Lu_prl06,Austin_sheath}, and the deflection of the bubble centroid by the injector pulse. Orthogonally polarized multi-cycle driver and NSC injector pulses co-propagate in tenuous plasma with $n_p=4.4\times 10^{18}/{\rm cm}^3$: see Fig.~\ref{fig:schematic} for a schematic, Table I for laser parameters, and SOM for simulations details. The injector pulse delay $\Delta T = 21 {\rm fs}$ is optimized to inject electrons near the back of the plasma bubble. Because of the low power and short duration of the injector pulse, its energy $U_{\rm inj} \sim 20{\rm mJ}$ is a small fraction of the driver pulse energy $U_{\rm dr} \sim 680{\rm mJ}$.

\begin{table}[ht]
\caption{\label{table:1}Parameters of the driver and injector pulses}
 \begin{ruledtabular}
 \begin{tabular}{lccc}
  Laser pulse& Driver & Injector 1 & Injector 2  \\
 \hline
  Polarization & y & z & z \\
  Wavelength& $\lambda_{\rm dr} = 0.8{\rm \mu m}$ & $\lambda_{\rm inj} = 2.4{\rm \mu m}$ & $\lambda_{\rm inj}^{(2)} = 1.2{\rm \mu m}$  \\
 FWHM & $T_{\rm dr}=22 {\rm fs}$ & $T_{\rm inj}= 6 {\rm fs}
 $ & $T_{\rm inj}^{(2)}= 3 {\rm fs}$\\
 Spot Size & $\sigma_{\rm dr} = 10 {\rm \mu m}$ &  $\sigma_{\rm inj} = 8 {\rm \mu m}$ & $\sigma_{\rm inj}^{(2)} = 8 {\rm \mu m}$\\
 Peak power & $P_{\rm dr}= 31 {\rm TW}$ & $P_{\rm inj}= 3.4 {\rm TW}$ & $P_{\rm inj}^{(2)}= 13.2 {\rm TW}$  \\
\end{tabular}
\end{ruledtabular}
\end{table}

During the early co-propagation stage of the driver and injector pulses (see Fig.~\ref{fig:evolution}(a)), the former produces the bubble while the latter induces its transverse centroid undulation in the injector polarization direction $z$. Bubble undulations manifest as a transverse on-axis wakefield $W_{\perp}\equiv E_z + B_y$ shown in Fig.~\ref{fig:evolution}(c), where $\left( E_z, B_y \right)(\zeta,z=y=0)$ are the transverse electric/magnetic bubble wakefields. In agreement with Eq.~(\ref{eq:period_cep}), the transverse wake oscillates with period $cT_{\rm CEP} \approx 70 \lambda_{\rm dr}$ before the injector pulse depletes around $x = L_{\rm inj}^{\rm depl} \approx 0.2 {\rm mm}$. The phase of $W_{\perp}$ is controlled by the injector CEP as shown in Fig.~ \ref{fig:evolution}(c), where the transverse wakes produced by the injector pulses with the initial values of $\phi_{\rm CEP}=0,\pi/2$ are phase-shifted by $90^{\circ}$ with respect to each other.

The key difference of the described two-pulse scenario with $\lambda_{\rm inj} \gg \lambda_{\rm dr}$ from the previously proposed scheme utilizing a single NSC injector pulse is that the driver pulse dramatically outruns the long-wavelength injector pulse. The latter rapidly depletes because of its stronger interaction with plasma: $n_{\rm crit}(\lambda_{\rm inj} \ll n_{\rm crit}(\lambda_{\rm dr}$. As can be seen from Fig.~\ref{fig:evolution}(b), the injector pulse decays after $L_{\rm inj}^{\rm depl}= 250 \lambda_{\rm dr}$ while the driver pulse remains unchanged. Injector pulse depletion is mirrored by the decay of the transverse plasma wake $W_{\perp}$ over the same distance $L_{\rm inj}^{\rm depl}$ marked by the blue dot-dashed line in Fig.~\ref{fig:evolution}(c). Therefore, the PUB injection is expected to stop after $x \approx L_{\rm inj}^{\rm depl} \ll L_{\rm acc}$, where $L_{\rm acc} \approx 1.5 {\rm mm}$ is the acceleration distance determined by the depletion of the driver. Because $L_{\rm acc} \gg L_{\rm inj}$, CEP-independent high-energy acceleration is expected.

PIC simulations confirm that bubble's centroid undulations induce electron injections into the bubble and generate micro-bunches shown in Fig.~\ref{fig:injected}(a). Several electron injections with regular spacing $T_{\rm CEP}/2 \approx 35 \lambda/c$ are shown in Fig.~\ref{fig:injected}(b). Such PUB-induced injection terminates after $x \approx L_{\rm inj}^{\rm depl}$. As the result, the total injected charge $Q_1\approx 93{\rm pC}$, is distributed over several micro-bunches, as shown in Fig.~\ref{fig:injected}(a,d). Electron density peaks are separated by $\Delta \xi \approx T_{\rm CEP}/4\gamma_b^2 \approx 0.66 \lambda_{\rm dr}$, as estimated earlier. Remarkably, the peak-current spike of $I\approx 20 {\rm kA}$ at $x-ct\approx 8\lambda_{\rm dr}$ contains $q\approx 16 {\rm pC}$ of charge compressed into a sub-femtosecond-scale time interval of $\delta \xi/c \approx 0.8 {\rm fs}$. One recently proposed mechanism for generating ultra-short current spikes utilizes sub-mm regions of laser-ionized high-$Z$~\cite{mori_bunches_prl16}. In contrast, the described PUB-based approach works for uniform fully-ionized plasmas. We also note that the PUB-based injection described here owes to transverse bubble undulations and not to the accompanying modulation of its overall size\cite{kalmykov,gonsalves_nphys11,austin_injection} due to rapid extinction of the NSC pulse. This was verified using 2D PIC simulations, where only one (in-plane) polarization of the NSC pulse resulted in bubble undulation and electron injection. The orthogonal polarization produced neither (see Fig.S6 of the SOM).

The two major current spikes marked as a square and a star in Fig.~\ref{fig:injected}(d) originate from the two correspondingly marked electron injections at $ct_1\approx 135\lambda_{\rm dr}$ and $ct_2\approx 250 \lambda_{\rm dr}$ shown in Fig.~\ref{fig:injected}(b). The current spikes correspond to quasi-monoenergetic electron bunches that develop after the propagation distance $L_{\rm acc}$; their prominent spectral peaks at $\gamma_1 \approx 630$ (square) and $\gamma_2 \approx 520$ (star) are plotted in Fig.~\ref{fig:injected}(c). From the inset showing electron energy spectra for $\phi_{\rm CEP}=0$ (solid line) and $\phi_{\rm CEP}=\pi/2$ (dashed line) around $\gamma_1$, we confirm that those are indeed nearly CEP-independent as the consequence of $L_{\rm acc} \gg L_{\rm inj}^{\rm depl}$. Therefore, it is possible to take advantage of CEP-controlled electron injection while maintaining near-independence of the electron energy spectra on $\phi_{\rm CEP}$.

Note that the electrons injected at the earlier time ($t_1 < t_2$) reach higher energy ($\gamma_1 > \gamma_2$) because they partially deplete the wakefield, reducing the acceleration gradient for subsequently injected electrons. Also noteworthy is a small energy peak at $\gamma_1 \approx 670$ corresponding to the earliest injection at $ct_0\approx 100 \lambda_{\rm dr}$. This injection contains fewer electrons than the next injection at $t=t_1$: significant bubble centroid oscillation after the first injection {\it de-traps} some of the earliest-injected electrons. Therefore, bubble undulations can, in principle, cause both electron trapping and de-trapping.

The described injection/acceleration approach utilizing a long-wavelength NSC injector was compared to the following alternative scenarios: (i) no injector pulse, (ii) short-wavelength ($\lambda_{\rm inj}^{(2)} = \lambda_{\rm inj}/2$) injector pulse with $P_{\rm inj}^{(2)} = 4P_{\rm inj} = 13.2 {\rm TW}$ and $\tau = 3{\rm fs}$. Note that while the injector pulse alone can also inject/accelerate electrons, it cannot sustain a stable accelerating bubble over a significant~\cite{Zhengyan, CEP_observable,Salehi} distance, resulting in electron energy gain of less than $10$ MeV and injected charge less than $q\approx 16{\rm pC}$. Scenario (i) yields a small accelerated charge $q\approx 1 {\rm pC}$, because the slowly-evolving driver pulse cannot efficiently inject electrons.

Scenario (ii), designed to preserve the injector pulse's ponderomotive potential $U_p \propto P_{\rm inj}\lambda_{\rm inj}^2/\sigma_{\rm inj}^2$ while reducing its wavelength, also results in inefficient charge injection. The corresponding current profile is indicated by a red line in Fig.~\ref{fig:injected}(d), and the total injected charge $Q_2\approx 21{\rm pC} \ll Q_1$ [Fig.~\ref{fig:injected}(c)].  Qualitatively, this is because the non-ponderomotive scaling of the injector-induced bubble undulation favors longer injector wavelengths as predicted by Fig.~\ref{fig:trajectory}(d). The smaller bubble undulation amplitude is further evidenced by the smaller $W_\perp$ amplitude (dashed line in Fig. ~\ref{fig:evolution}(d) for $\lambda_{\rm inj}^{(2)}$) compared to the $\lambda_{\rm inj}$ case (solid line).

In conclusion, we propose and theoretically demonstrate a two-pulse CEP-controlled scheme for injecting and accelerating electrons from a preformed plasma. By combining a near-single-cycle long-wavelength laser pulse for rapid electron injection with a multi-cycle short-wavelength driver pulse for long-distance acceleration of the injected electrons, we demonstrate that sub-femtosecond high-current (tens of kA) electron micro-bunches with ultra-relativistic energies (hundreds of MeVs) can be generated. We conjecture that such beams can be used as a compact source of ultra-short X-ray radiation owing to their high energy and large-amplitude betatron motion.

 \section{Acknowledgments}
This work was supported by the DOE Grant No. DE-SC-0019431. The authors thank the Texas Advanced Computing Center (TACC) at The University of Texas at Austin for providing the HPC resources.

\end{document}